# Evidence of a Hybridized Topological State in Weyl Semimetal/Topological Insulator $Mn_{3+x}Sn_{1-x}/Bi_{0.85}Sb_{0.15}$ Heterostructures

*Ryan T. Van Haren[1*] (corresponding author), Gillian P Boyce[1], Gregory M. Stephen[1], Vinay Sharma[1], Don Heiman[2], Aubrey T. Hanbicki[1], and Adam L. Friedman[1]*

Email of corresponding author: **rvanhare@umd.edu**

[1] Laboratory for Physical Sciences, 8050 Greenmead Dr., College Park, MD, 20740 USA

[2] Department of Physics, Northeastern University, Boston, MA 02115, USA

**Abstract**

We report magnetotransport evidence of a hybridized Weyl semimetal (WSM) Fermi arc/topological insulator (TI) surface state at the interface of a ferromagnetic $Mn_{3+x}Sn_{1-x}/Bi_{0.85}Sb_{0.15}$ heterostructure. High target utilization sputtering (HiTUS) was used to grow polycrystalline $Mn_{3+x}Sn_{1-x}$ films and $Mn_{3+x}Sn_{1-x}/Bi_{0.85}Sb_{0.15}$ heterostructures on thermally oxidized $Si/SiO_2$ (100) substrates that exhibit the negative coefficient anomalous Hall effect (AHE) resulting from topological Weyl node transport. When various defects and impurities are introduced into these $Mn_{3+x}Sn_{1-x}$ films, a ferromagnetic (FM) phase develops that practically eliminates the topological Weyl node conduction. These FM $Mn_{3+x}Sn_{1-x}$ films exhibit large exchange bias effects below $T = 200$ K that we attribute to the coexistence of a FM phase and the triangular antiferromagnetic (AFM) WSM phase. When $Bi_{0.85}Sb_{0.15}$ overlayers are grown on the FM $Mn_{3+x}Sn_{1-x}$ films, the magnetotransport signal of Weyl node topological transport is restored, an effect we do not observe when replacing the $Bi_{0.85}Sb_{0.15}$ TI with heavy metal overlayers. We attribute the restoration of the Weyl node topological transport to the formation of a hybridized topological state at the WSM/TI interface.

**Introduction**

Motivated by slowing CMOS performance gains, desperate need for increased efficiency for the AI boom, and new device paradigms required to support the quantum computing ecosystem, a renewed interest in unconventional magnetic materials could realize solution pathways for next-generation technologies[1–3]. Among these unconventional magnetic materials is the room temperature antiferromagnetic Weyl semimetal (WSM) $Mn_3Sn$. Antiferromagnetic (AFM) materials may offer advantages over ferromagnetic (FM) materials in memory devices due to an absence of stray fields, thereby allowing increased densities, and higher operational speeds due to a greater magnetic resonance frequency[4–6].

At room temperature, the $Mn_3Sn$ hexagonal crystal orders into a non-collinear, triangular AFM phase[7–9]. Magnetotransport measurements in this phase exhibit a large anomalous Hall effect (AHE) with a negative coefficient, the result of the non-trivial Berry curvature acquired by the electron wavefunction during topological Weyl node transport in the WSM. This negative coefficient AHE contrasts with the positive anomalous Hall coefficient of ordinary FMs[10,11]. Below room temperature, $Mn_3Sn$ crystals can undergo several magnetic phase transitions: first, from the triangular AFM phase to a helical or incommensurate AFM phase near $T = 200$ K, then to a spin glass phase near $T = 10$ K[8,9,15–1715,16,18,19]

Previous work on $Mn_3Sn$ films have attempted to harness the unique magnetic and spin properties of this crystal through the growth of $Mn_3Sn$/heavy metal bilayer spin-orbit torque (SOT) devices[20,21]. In these experimental $Mn_3Sn$/heavy metal SOT devices, a large spin-orbit coupling (SOC) is a result of the large atoms of the heavy metal. However, a topological insulator (TI), like $Bi_{1-x}Sb_x$, is expected to generate much larger spin currents at these types of interfaces due to its spin-momentum locked topological surface states, creating opportunities for drastically increased device efficiency[22–26]. While many WSMs are relatively fragile materials and fabrication of novel heterostructures is difficult, $Mn_3Sn$ is robust to sputtered metallic overlayers, therefore it is an excellent candidate with which to study WSM/TI interfaces. However, the growth and characterization of such a heterostructure have not previously been described.

Here, we present the growth and characterization of a WSM/TI heterostructure. Using high target utilization sputtering (HiTUS), we deposit polycrystalline WSM $Mn_{3+x}Sn_{1-x}$/$Bi_{0.85}Sb_{0.15}$ (003) layers on thermally oxidized Si/$SiO_2$ (100) substrates. Magnetotransport measurements of these heterostructures reveal a negative coefficient AHE at room temperature indicative of Weyl node transport, thus confirming the WSM nature of HiTUS-grown $Mn_{3+x}Sn_{1-x}$/$Bi_{0.85}Sb_{0.15}$ heterostructures. Second, we find that when various impurities or dopant layers are added to the $Mn_{3+x}Sn_{1-x}$ films during growth, Weyl node transport is entirely suppressed, with the negative coefficient AHE characteristic of WSM $Mn_{3+x}Sn_{1-x}$ replaced by a FM, positive coefficient AHE. Finally, we recreate the $Mn_{3+x}Sn_{1-x}$/$Bi_{0.85}Sb_{0.15}$ heterostructure with these FM $Mn_{3+x}Sn_{1-x}$ films. Unlike the first WSM heterostructures, which displayed negative coefficient AHE at room temperature, these FM $Mn_{3+x}Sn_{1-x}$/$Bi_{0.85}Sb_{0.15}$ heterostructures display asymmetric "horns” in magnetotransport, with both positive and negative coefficient contributions to the AHE. We demonstrate that this behavior is result of a hybridized WSM/TI topological state that restores Weyl node transport.

**Methods**

We deposited thin film $Mn_{3+x}Sn_{1-x}/Bi_{0.85}Sb_{0.15}$ heterostructures via a high target utilization sputtering (HiTUS) system (PlasmaQuest/QuanVerge) onto thermally oxidized $Si/SiO_2$ (100) substrates from stoichiometric $Mn_3Sn$ and $Bi_{0.85}Sb_{0.15}$ sputtering targets. A diagram of the $Mn_{3+x}Sn_{1-x}/Bi_{0.85}Sb_{0.15}$ heterostructure stack is shown in Figure 1(a). The HiTUS system creates a plasma remotely, and thus decouples the plasma density from the sputtering energy, opening up additional growth parameter space and resulting in improved crystallinity in deposited films as compared to conventional magnetron sputtering[33,34]. First, we deposited $Mn_{3+x}Sn_{1-x}$ directly onto $Si/SiO_2$ substrates at ambient temperature using an Ar plasma. After deposition, we annealed the samples in vacuum at $T = 450$℃ for 1 hour. The samples were allowed to cool back to ambient temperature in vacuum and then we deposited the $Bi_{0.85}Sb_{0.15}$ film, with growth parameters described in a previous work[34]. After deposition of the $Bi_{0.85}Sb_{0.15}$ film, we annealed the heterostructure stacks in vacuum at $T = 175$℃ for 1 hour to crystallize the $Bi_{0.85}Sb_{0.15}$ film. After cooling back to ambient temperature, the stack was capped with a thin MgO film to protect the stack from atmosphere.

Dopants were introduced into the $Mn_{3+x}Sn_{1-x}$ films by two methods. First, fluorine was unintentionally introduced into the early $Mn_{3+x}Sn_{1-x}$ films during growth due to impurities in the sputtering target. Second, a thin dopant layer was introduced into the $Mn_{3+x}Sn_{1-x}$ films by interrupting the deposition of the $Mn_{3+x}Sn_{1-x}$ film halfway through to deposit a layer of either Pt or $Ni_{81}Fe_{19}$. The remaining $Mn_{3+x}Sn_{1-x}$ film was then deposited, and the entire stack was annealed in the same way as the other films.

We fabricated Hall bar devices with a central channel 1.2 mm long and 0.16 mm wide using a shadow mask during sputtering deposition of the film, with one of the Hall bar devices shown in Figure 1(b). Magnetotransport measurements were performed in a 1 T electromagnet with a 4 K closed-cycle cryostat using standard DC source-measure units, and in a Quantum Design Physical Property Measurement System (PPMS). Magnetic moment measurements of the films were performed on a SQUID (superconducting quantum interference device) magnetometer. First order reversal curve (FORC) analysis of minor loop measurements was executed using custom MATLAB code, as described elsewhere[35].

**Results and Discussion**

The magnetotransport characteristics of polycrystalline WSM $Mn_{3+x}Sn_{1-x}$ thin films are well demonstrated by the AHE loop of our HiTUS-grown WSM $Mn_{3+x}Sn_{1-x}/Bi_{0.85}Sb_{0.15}$ heterostructure, shown in Figure 1(c), and the AHE loop of bare WSM $Mn_{3+x}Sn_{1-x}$, shown in Figure 2(a). At $T = 300$ K, in both the heterostructure and bare film we observe a negative coefficient AHE indicative of topological Weyl node transport and characteristic of WSM phase $Mn_{3+x}Sn_{1-x}$[12–14]. As the WSM $Mn_{3+x}Sn_{1-x}/Bi_{0.85}Sb_{0.15}$ heterostructure is cooled, the negative AHE signal decreases in magnitude and eventually flips sign below $T = 200$ K, as shown in Figure 1(c), when the $Mn_{3+x}Sn_{1-x}$ transitions from the WSM, triangular AFM phase to an incommensurate magnetic phase with weak ferromagnetism and no Weyl node transport[15,16].

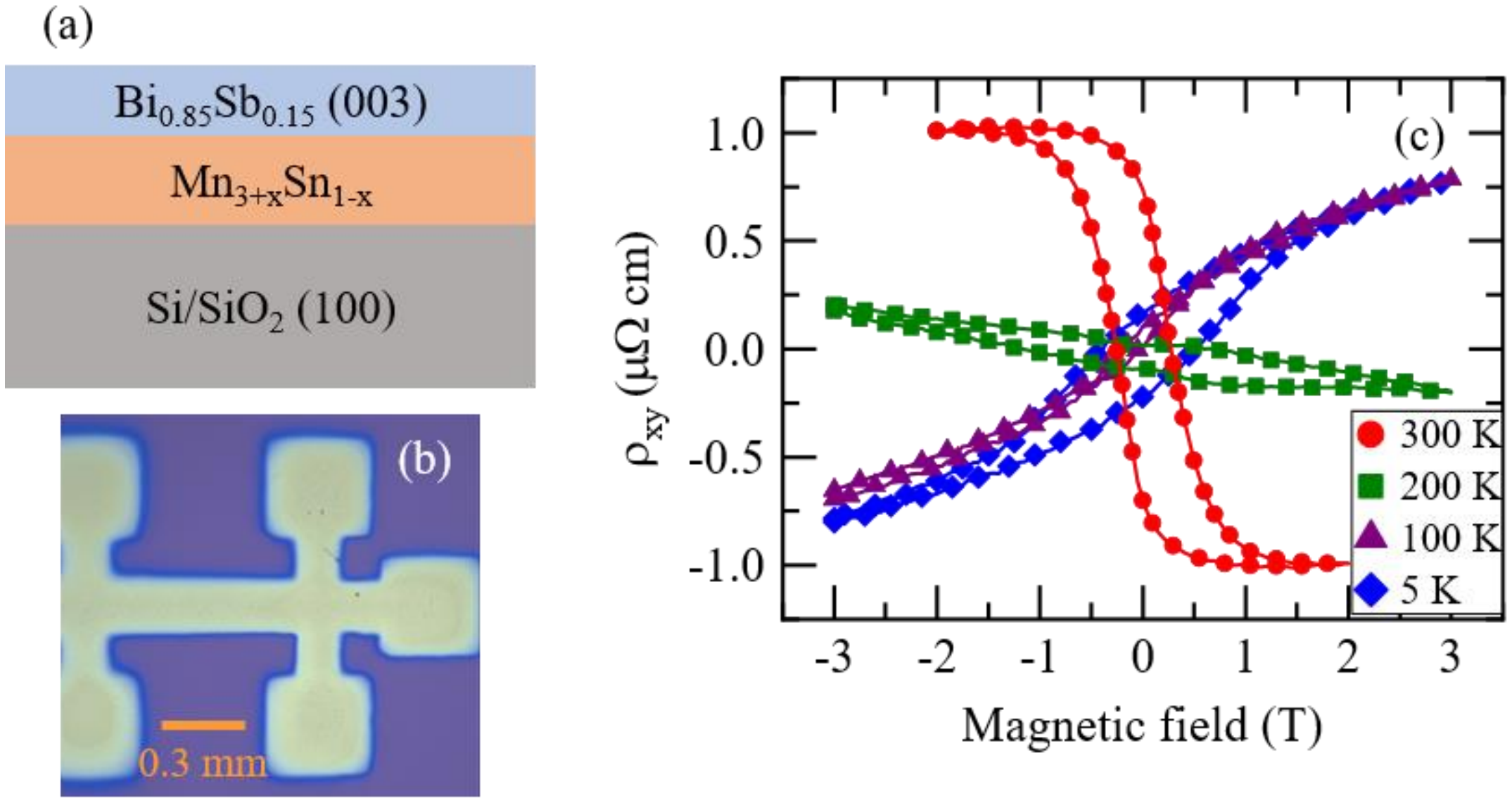

**Figure 1: (a) Schematic of $Mn_{3+x}Sn_{1-x}/Bi_{0.85}Sb_{0.15}$ heterostructure stack sequence. (b) Micrograph of processed Hall Bar device. (c) Temperature dependence of WSM $Mn_{3+x}Sn_{1-x}/Bi_{0.85}Sb_{0.15}$ heterostructure Hall resistivity, demonstrating the phase change of the $Mn_{3+x}Sn_{1-x}$ crystal from the WSM triangular AFM phase at room temperature, characterized by a negative coefficient AHE, to a weak FM incommensurate phase with positive coefficient AHE below 200 K.**

By introducing dopants and impurities into our $Mn_{3+x}Sn_{1-x}$ thin film growths, we can virtually eliminate the magnetotransport signal of Weyl node transport at room temperature. Figure 2(b) shows the magnetoresistivity of a $Mn_{3+x}Sn_{1-x}$ film with fluorine impurities, resulting in a positive coefficient AHE as expected for an ordinary FM. These FM $Mn_{3+x}Sn_{1-x}$ films retain the same polycrystalline structure as un-doped, WSM $Mn_{3+x}Sn_{1-x}$ films, evidenced by X-Ray diffraction (XRD) spectra shown in the supplemental Fig. S1. We can also induce this transition from WSM to FM transport behavior in $Mn_{3+x}Sn_{1-x}$ films by introducing Pt or $Ni_{81}Fe_{19}$ doping layers into the $Mn_{3+x}Sn_{1-x}$ films during growth. AHE loops of these Pt or $Ni_{81}Fe_{19}$ doped films are shown in the supplemental Fig. S2. The apparent effect of these dopants in magnetotransport of the $Mn_{3+x}Sn_{1-x}$ films is the formation of a ferromagnetic phase that disrupts the Weyl node transport that would ordinarily occur. Weyl node transport is likely disrupted by either the opening of energy gaps at the Weyl nodes, or through dephasing events that decohere the Berry curvature of the electron wavefunction at a Weyl node.

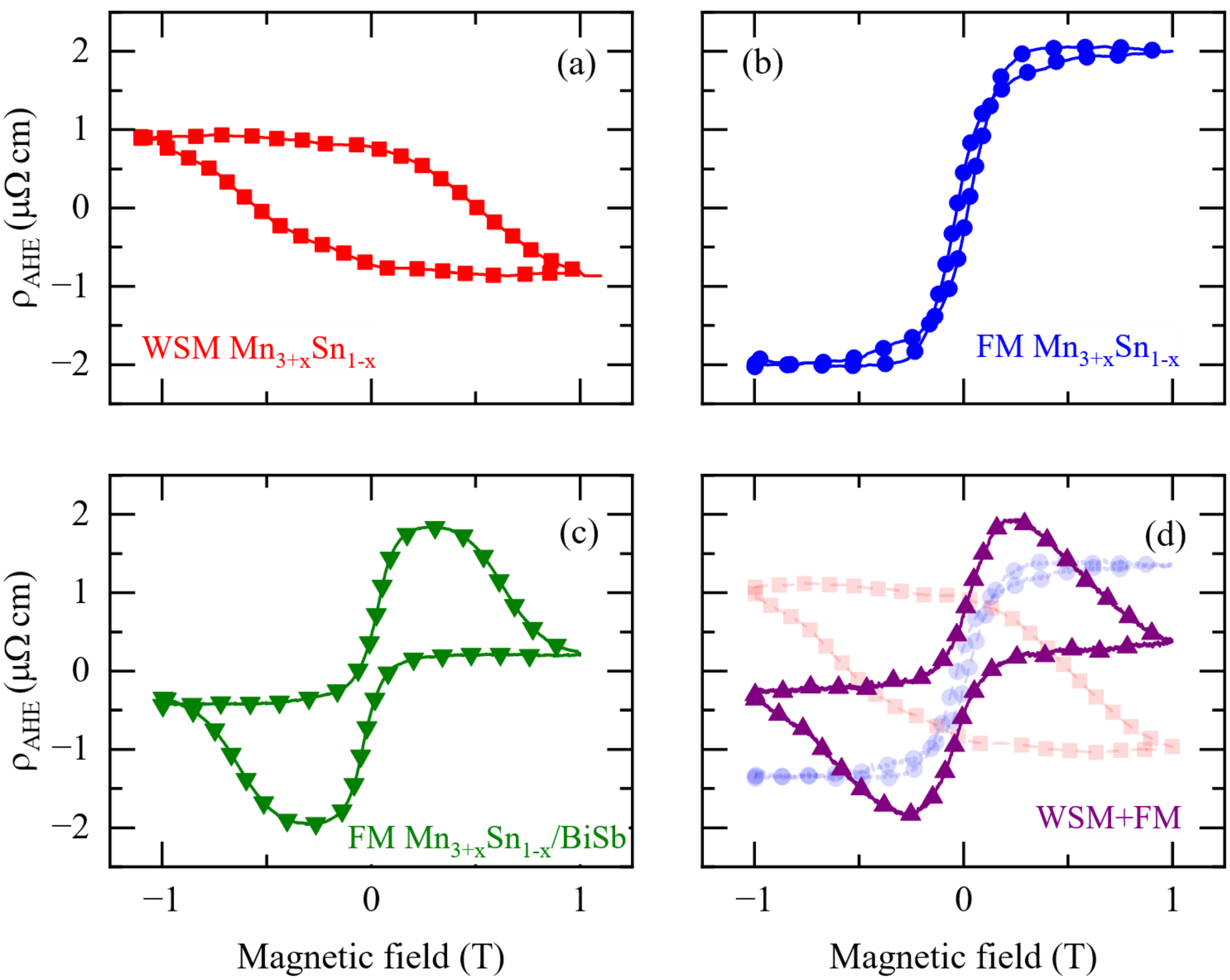

**Figure 2: Room temperature AHE measurements of (a) WSM $Mn_{3+x}Sn_{1-x}$ thin film exhibiting the negative AHE coefficient signal characteristic of Weyl node transport, (b) FM $Mn_{3+x}Sn_{1-x}$ thin film exhibiting the positive AHE coefficient signal characteristic of ferromagnets, and (c) FM $Mn_{3+x}Sn_{1-x}/Bi_{0.85}Sb_{0.15}$ heterostructure that exhibits magnetotransport "horns." (d) AHE signal resulting from the superposition of (a) and (b), indicated by the transparent curves, exhibiting striking qualitative similarity to the AHE signal of (c).**

Growing a $Bi_{0.85}Sb_{0.15}$ TI overlayer on FM $Mn_{3+x}Sn_{1-x}$ films significantly alters the magnetotransport behavior. As shown in Figure 2(c), large asymmetric "horns" develop in the AHE loop. These horns are reminiscent of a topological Hall effect (THE) that manifests as a result of a chiral spin texture in thin films. Reports of a THE in $Mn_3Sn$/heavy-metal bilayer films resulting from chiral spin textures have been made previously, but the behavior observed in our samples differs from previous reports in film composition, magnitude of the effect, and the temperature region over which it occurs[36–39]. While we were able to reproduce the magnetotransport horns across several FM $Mn_{3+x}Sn_{1-x}/Bi_{0.85}Sb_{0.15}$ samples, we did not observe any such effect in FM $Mn_{3+x}Sn_{1-x}$ films with other strong SOC overlayers, such as Pt or Ta, as shown in supplemental Fig. S2, ruling out ordinary SOC effects or interface roughness as possible explanations.

Although this asymmetric AHE signal appears superficially like a THE, conclusively determining the presence of a chiral spin texture from magnetotransport alone requires careful consideration. Nearly identical transport behavior can result from the superposition of two AHE curves with opposite AHE coefficients and different coercivities[35,40–43]. An example of this AHE superposition is shown in Figure 2(d), which is the sum of the AHE signals from the WSM and FM $Mn_{3+x}Sn_{1-x}$ samples shown in Figure 2(a) and (b). This sum qualitatively recreates the AHE signal measured in the FM $Mn_{3+x}Sn_{1-x}/Bi_{0.85}Sb_{0.15}$ heterostructure. This will be discussed in more detail below.

Minor loop FORC measurements, where the applied field is swept from a reversal field with magnitude less than magnetic saturation, are a powerful technique for identifying the origin of magnetotransport phenomena[35,42]. By taking a series of magnetoresistivity measurements across different reversal fields, the magnetic switching dynamics of a sample can be mapped. Because a THE is not a magnetic transition and has no net magnetization outside of the horn, its behavior in a minor loop measurement will appear differently than two superimposed AHE that switch at different applied fields[35,40–42]. Even very subtle THE or small superimposed AHE can be detected from magnetic minor loop measurements by using FORC analysis[35]. In FORC analysis, the partial derivative of the magnetoresistance of the minor loop is taken with respect to the applied magnetic field $H_a$ and the reversal field $H_r$,

$$\rho_{FORC}(H_a, H_r) = \frac{\partial^2 R}{\partial H_a \partial H_r}.$$

In this derivative, each magnetic transition with the "S" shape of a tanh function, will show up as a single peak in $\rho_{FORC}$, while the peaked THE signal will show up in $\rho_{FORC}$ as a pair of peaks with opposite signs. Plotted across a series of minor loops, $\rho_{FORC}$ forms a heat map as a function of $H_a$ and $H_r$, which can be transformed into the more intuitive coercive field $H_c = (H_a - H_r)/2$ and interaction or exchange field $H_u = (H_a + H_r)/2$. In the case of a THE superimposed on an AHE, we expect the heat map of $\rho_{FORC}$ to exhibit three peaks: two, coupled peaks split in $H_u$ and at the same value of $H_c$, corresponding to a THE and a third, unrelated peak corresponding to an AHE.

Room temperature minor loop magnetoresistance measurements of one of our FM $Mn_{3+x}Sn_{1-x}/Bi_{0.85}Sb_{0.15}$ heterostructures are shown in Figure 3(a). The second partial derivative of this minor loop series, plotted as a $\rho_{FORC}$ heat map as a function of $H_c$ and $H_u$, is shown in Figure 3(b). This second derivative exhibits only two, isolated peaks near ($H_c$, $H_u$) = (0 T, 0 T) and (0.5 T, 0 T), respectively, corresponding to two, incommensurate AHE and not a THE.

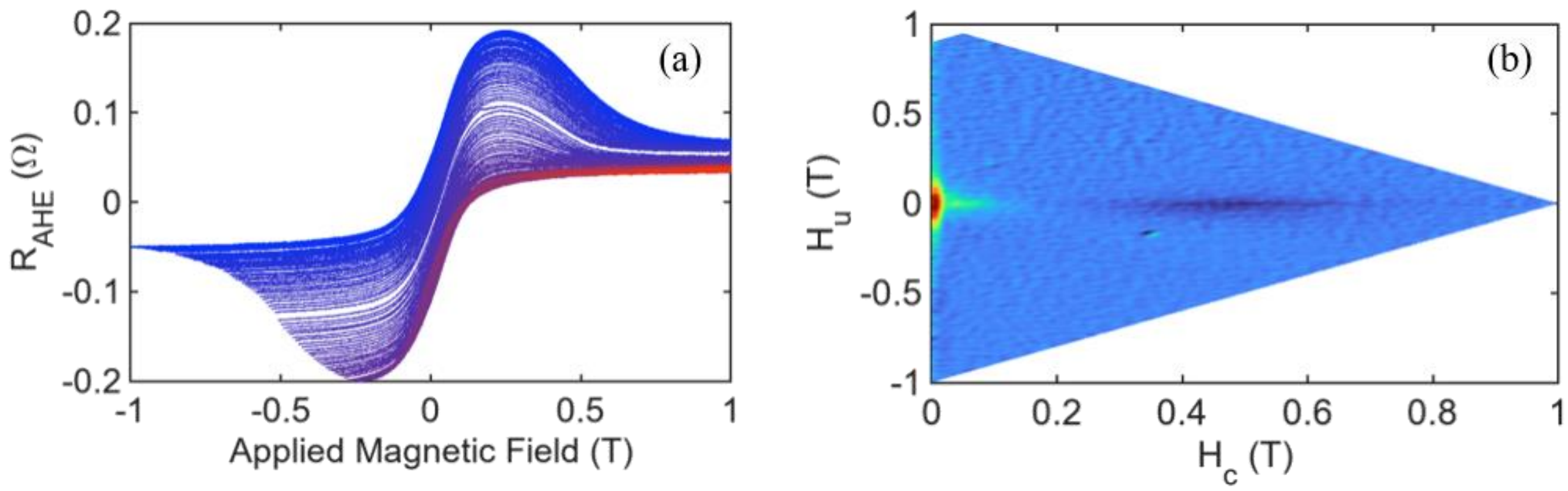

**Figure 3: (a) Minor loop series of a FM $Mn_{3+x}Sn_{1-x}/Bi_{0.85}Sb_{0.15}$ heterostructure at $T = 300$ K. (b) Heat map plot of $\rho_{FORC}$ derived from the same minor loop series, showing two peaks along $H_u$ = 0 corresponding to two AHE loops. The dimple near $(H_u, H_c) = (0.2, 0.35)$ T is an artifact due to an irreproducible change in the magnetoresistance curve of applied magnetic field.**

FORC analysis of magnetoresistance indicates no chiral spin texture, and we cannot replicate this magnetotransport behavior with large SOC heavy metals. Therefore, we propose that the magnetotransport horns observed in our FM $Mn_{3+x}Sn_{1-x}/Bi_{0.85}Sb_{0.15}$ heterostructures are evidence of a hybridized topological state that forms at the WSM/TI interface, resulting in two conducting channels that each contribute an AHE component: a FM channel from conduction through the bulk of the FM $Mn_{3+x}Sn_{1-x}$, and a restored Weyl node transport channel at the $Bi_{0.85}Sb_{0.15}$ interface. Note again, that Figure 2(d), is a simple superposition of the WSM and FM $Mn_{3+x}Sn_{1-x}$ AHE signals, and this simple combination recreates the magnetotransport horns in the FM $Mn_{3+x}Sn_{1-x}/Bi_{0.85}Sb_{0.15}$ heterostructures almost perfectly with no fitting parameters.

Evidently, the introduction of the $Bi_{0.85}Sb_{0.15}$ overlayer restores Weyl node transport in the FM $Mn_{3+x}Sn_{1-x}$ film, creating a conducting channel that exhibits the negative coefficient AHE characteristic of the WSM state. This conducting channel is likely topological in nature, as growing identical heterostructures with strong SOC metals that lack topological phases, like Ta and Pt, exhibited only the FM $Mn_{3+x}Sn_{1-x}$ transport behavior. A WSM/TI hybridized state, similar to what we propose, has been previously predicted to occur at the interface between of a WSM and a TI[30], and stems from the fact that the topological 3D and 2D states of TIs and WSMs, respectively, are mathematically and topologically closely related to each other[27–32].

The temperature dependence of the AHE loops of FM $Mn_{3+x}Sn_{1-x}$ and FM $Mn_{3+x}Sn_{1-x}/Bi_{0.85}Sb_{0.15}$ heterostructures provide supporting evidence for the existence of a hybridized topological conducting channel in the heterostructures. As shown in Figure 4(a), a striking feature of the two-channel AHE observed in the FM $Mn_{3+x}Sn_{1-x}/Bi_{0.85}Sb_{0.15}$ heterostructures is its persistence from $T = 300$ K to $T = 5$ K. As described previously, the negative coefficient AHE component is an indication of Weyl node transport that occurs when the $Mn_{3+x}Sn_{1-x}$ is in the triangular AFM phase. This phase of $Mn_{3+x}Sn_{1-x}$ can be stabilized down to low temperature with film stoichiometry[16]. Our observation of a negative coefficient AHE component down to $T = 5$ K implies stabilization of the triangular AFM phase in these films, alongside a FM phase that results in the positive coefficient AHE.

Coexistence of the triangular AFM and FM phases in FM $Mn_{3+x}Sn_{1-x}$ is further evidenced by observations of magnetic exchange bias in our FM $Mn_{3+x}Sn_{1-x}$ films and heterostructures[44,45]. Exchange bias in the magnetotransport signal of a bare FM $Mn_{3+x}Sn_{1-x}$ film is shown in Figure 4(b). It first appears at $T = 200$ K and persists down to $T = 5$ K. The appearance of exchange bias in the bare film indicates domains of FM and AFM order with exchange bias coupling at the interfaces between these domains[46]. Nearly identical exchange bias behavior is observed in the magnetic moment of FM $Mn_{3+x}Sn_{1-x}/Bi_{0.85}Sb_{0.15}$ heterostructures, as shown in Figure 4(c). The direction of the shift can be controlled by switching the polarity of the field during the cool down, as shown in Figure 4(d), indicating a genuine exchange bias effect. These measurements demonstrate how, even though the triangular AFM WSM phase is stabilized in our FM $Mn_{3+x}Sn_{1-}$ $_x$ films down to cryogenic temperatures, the Weyl node transport associated with this phase is suppressed by coexisting FM domains. By growing a $Bi_{0.85}Sb_{0.15}$ TI overlayer, a hybridized topological channel is created that restores Weyl node transport and manifests as the asymmetric magnetotransport signal.

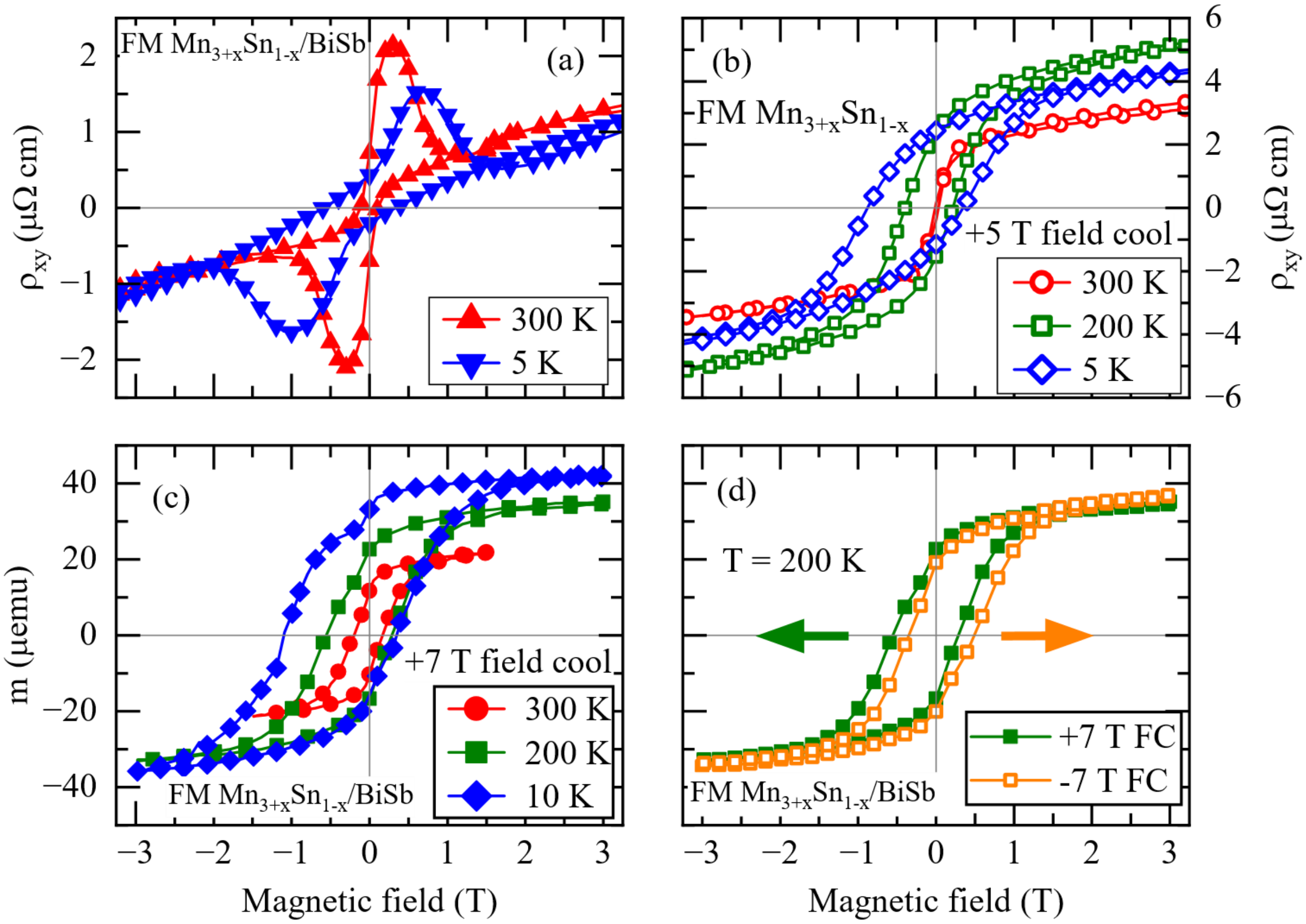

**Figure 4: (a) Hall resistivity of FM $Mn_{3+x}Sn_{1-x}/Bi_{0.85}Sb_{0.15}$ heterostructure, exhibiting asymmetric horns at $T = 300$ K and $T = 5$ K. (b) Hall resistivity of a FM $Mn_{3+x}Sn_{1-x}$ film, exhibiting exchange bias that develops below $T = 200$ K. (c) Magnetic moment measurement of FM $Mn_{3+x}Sn_{1-x}$/ $Bi_{0.85}Sb_{0.15}$ heterostructure also exhibiting exchange bias below $T = 200$ K. (d) Exchange bias effect in FM $Mn_{3+x}Sn_{1-x}$/ $Bi_{0.85}Sb_{0.15}$ shifting left and right depending on the polarity of the cooling field.**

**Conclusion**

In this work we have presented measurements of sputter deposited polycrystalline WSM $Mn_{3+x}Sn_{1-x}$ films with Weyl node transport and defected, polycrystalline FM $Mn_{3+x}Sn_{1-x}$ films with distinctly different properties from those previously studied. In these defect-added FM $Mn_{3+x}Sn_{1-x}$ films, the topological Weyl node transport that results in a negative coefficient AHE is suppressed by a ferromagnetic moment. When a $Bi_{0.85}Sb_{0.15}$ TI overlayer is grown on these FM $Mn_{3+x}Sn_{1-x}$ films, Weyl node transport is restored, resulting in the formation of asymmetric horns in magnetotransport. Minor loop FORC analysis indicates that the cause of the asymmetric magnetotransport signal is two superimposed AHE loops with opposite polarity and not a THE from a chiral spin texture. This two-channel, two-AHE interpretation is supported by magnetotransport and magnetic measurements of exchange bias in these FM $Mn_{3+x}Sn_{1-x}/Bi_{0.85}Sb_{0.15}$ heterostructures, which demonstrate that the necessary triangular AFM WSM phase of $Mn_{3+x}Sn_{1-x}$ is stabilized down to cryogenic temperatures. The addition of the $Bi_{0.85}Sb_{0.15}$ TI overlayer creates a hybridized topological conducting channel that restores Weyl node transport, with the asymmetric magnetotransport signal representing a superposition of both the negative coefficient AHE from Weyl node transport in the hybridized channel and the positive coefficient AHE from the FM $Mn_{3+x}Sn_{1-x}$ bulk channel. These measurements and analysis indicate the rich and unexplored topological phase space that can be accessed in WSM/TI heterostructures and motivates further work to study these hybridized topological states and their exotic properties.

ACKNOWLEDGMENT

The authors from LPS gratefully acknowledge critical assistance from LPS support staff, including G. Latini, J. Wood, K. Kim, R. Brun, P. Davis, and D. Crouse.

**Supplemental Material**

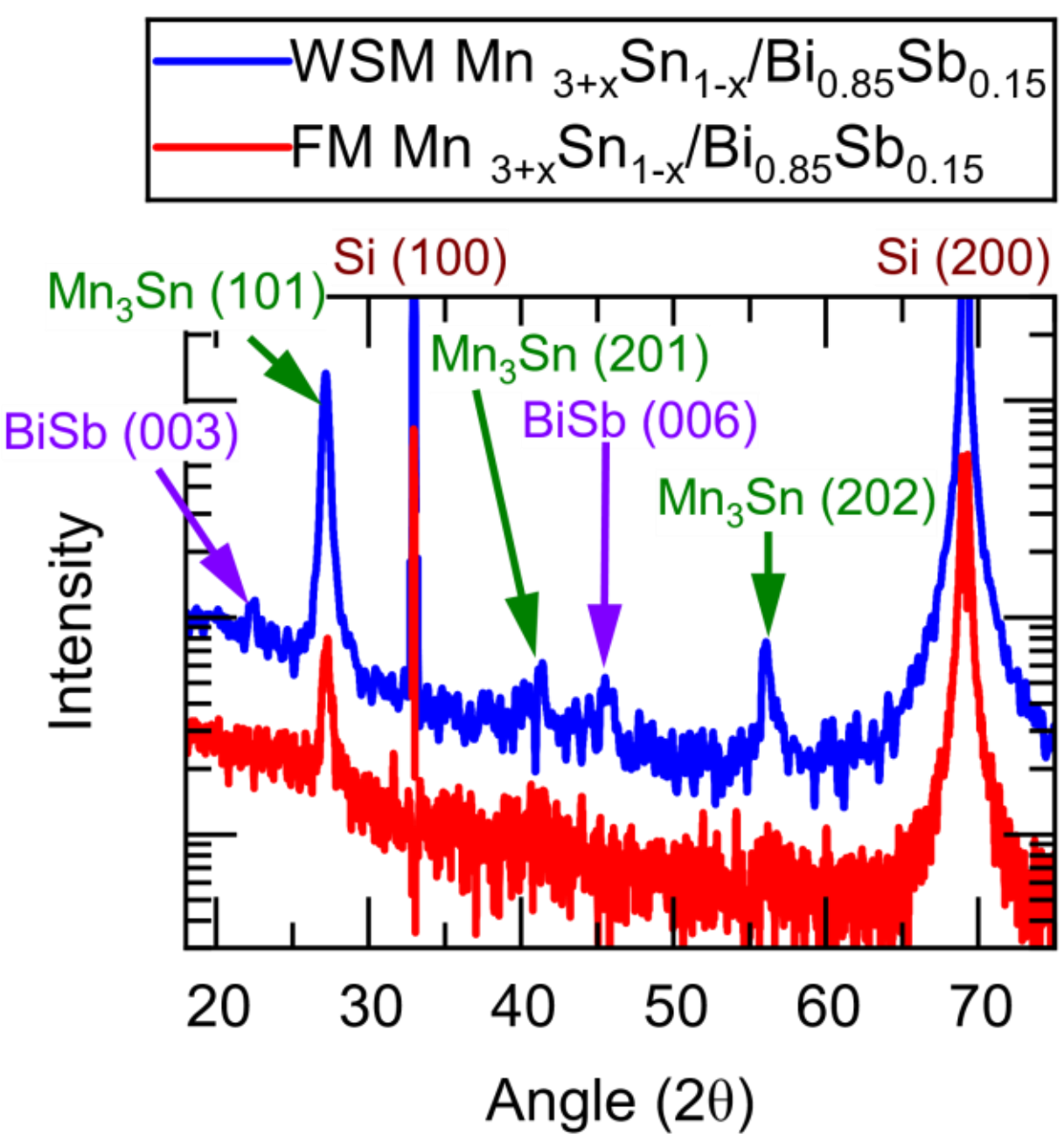

Figure S5: X-ray diffraction pattern (XRD) of two $Mn_{3+x}Sn_{1-x}/Bi_{0.85}Sb_{0.15}$ heterostructures showing the {101} peaks of $Mn_3Sn$ and the {003} peaks of $Bi_{0.85}Sb_{0.15}$. The top pattern is the WSM $Mn_{3+x}Sn_{1-x}/Bi_{0.85}Sb_{0.15}$ heterostructure that exhibits Weyl node transport and no magnetotransport horns. The bottom pattern is the FM $Mn_{3+x}Sn_{1-x}/Bi_{0.85}Sb_{0.15}$ heterostructure that exhibits magnetotransport horns and no Weyl node transport. In both samples, the {101} peaks are shifted slightly to the right from their expected positions for $Mn_3Sn$. This shift corresponds to a reduced lattice constant and slightly Mn rich $Mn_{3+x}Sn_{1-x}$ stoichiometry, in agreement with Ref 15.

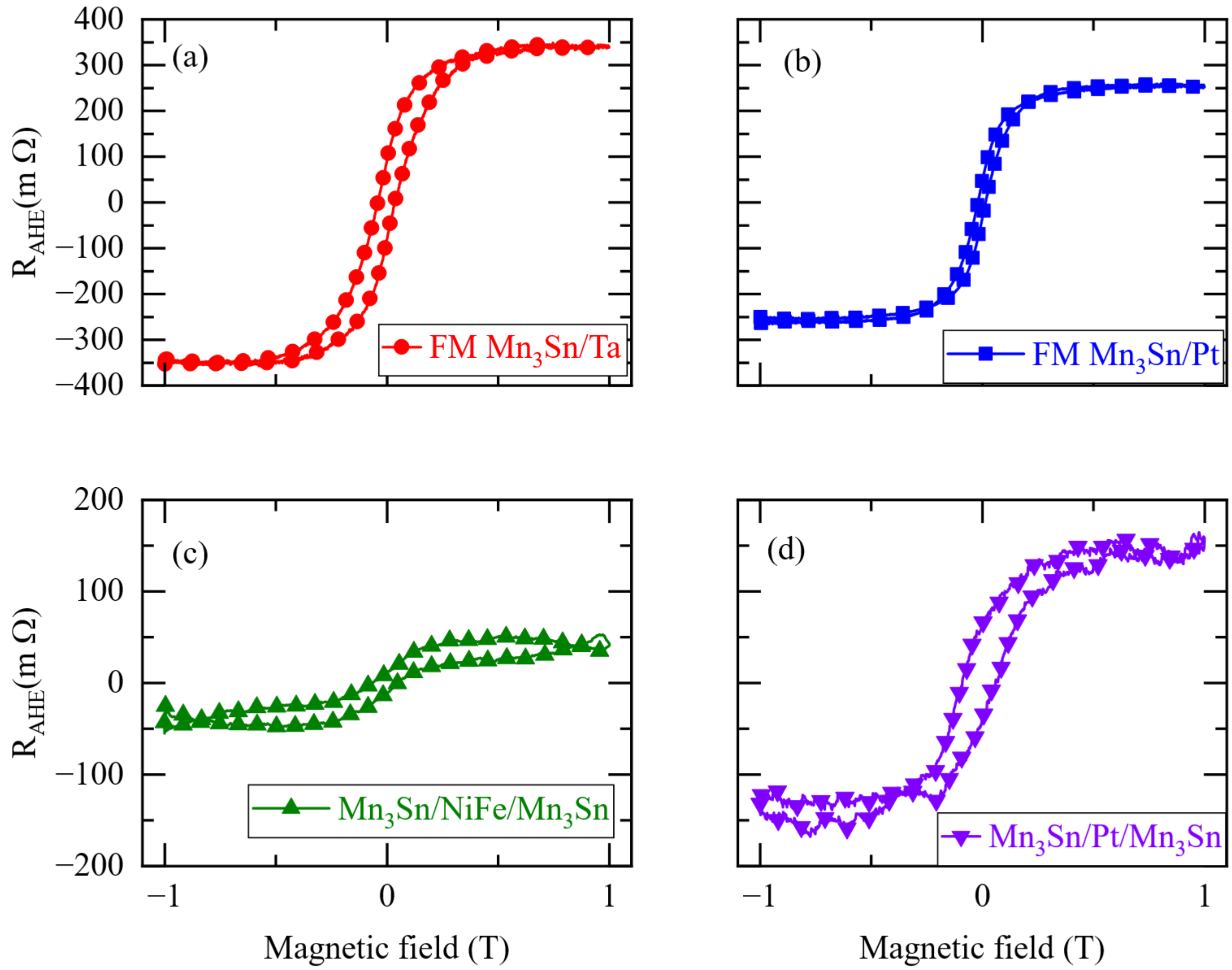

Figure S6: Electrical resistance as a function of applied magnetic field of various $Mn_{3+x}Sn_{1-x}$ films at room temperature. (a) Electrical resistance of a FM $Mn_{3+x}Sn_{1-x}$ film capped with Ta, exhibiting ferromagnetic AHE but no magnetotransport horns. (b) Electrical resistance of FM $Mn_{3+x}Sn_{1-x}$ capped with Pt, exhibiting the same behavior. (c) Electrical resistance of a WSM $Mn_{3+x}Sn_{1-x}$ film with a thin $Ni_{0.81}Fe_{0.19}$ layer introduced into the middle of the $Mn_{3+x}Sn_{1-x}$ film during growth, which switches the polarity of the WSM $Mn_{3+x}Sn_{1-x}$ AHE. (d) Electrical resistance of a WSM $Mn_{3+x}Sn_{1-x}$ film with a thin layer of Pt introduced into the middle of the $Mn_{3+x}Sn_{1-x}$ film in the same manner, exhibiting a similar AHE polarity flip.